\documentclass[a4paper,nopacs,12pt]{iopart}
\usepackage{amssymb,epsfig,bm}
\usepackage{graphicx,iopams,setstack}
\usepackage{psfrag,textcomp}
\usepackage{upgreek,cite}

\newcommand{\bracket}[1]{\left\langle #1\right\rangle}

\newcommand{\beeq}[1] {\begin{eqnarray}#1\end{eqnarray}}
\newcommand{\beeqn}[1] {\begin{eqnarray*}#1\nonumber\end{eqnarray*}}

\begin{document}
\title[]{A unified approach for large deviations of bulk and extreme eigenvalues of the Wishart ensemble}
\author{Adolfo Camacho Melo}
\address{Facultad de ciencias, UNAM, Avenida Universidad 3000, Circuito Exterior S/N, Coyoac\'an, Cd. Universitaria, 04510   M\'exico D.F., M\'exico}
\author{Isaac P\'erez Castillo}
\address{Departamento de Sistemas Complejos, Instituto de F\'isica, UNAM, P.O. Box 20-364, 01000 M\'exico D.F., M\'exico}
\begin{abstract}
Within the framework of the Coulomb fluid picture, we present a unified approach to derive the large deviations of bulk and extreme eigenvalues of large Wishart matrices. By analysing the statistics of the shifted index number  we are able to derive a rate function $\Psi (c, x)$ depending on two variables:  the fraction $c$ of eigenvalues to the left of an infinite energetic barrier at position $x$. For  a fixed value of $c$, the rate function gives the large deviations of the bulk eigenvalues. In particular, in the limits $c\to 0$ or $c\to 1$ it is possible to extract the left and right deviations of the smallest and largest eigenvalues, respectively. Alternatively, for a fixed value $x$ of the barrier, the rate function provides the large deviations of the shifted index number.  All our  analytical findings are compared with Metropolis Monte Carlo simulations, obtaining excellent agreement. 
\end{abstract}

\maketitle
\section{Introduction}
With the advent of  modern technology many branches of science are literally soaked in humungous amounts of data,  some of it yet to be analysed and fully understood. How all this big data will be transformed into useful data depends primarily on the mathematical approaches at our disposal to analyse it. Consider, for instance, that the data collected from our system of interest is given by an  $N\times T$  matrix $X$ with entries $X_{it}$ with $i=1,\ldots,N$ and $t=1,\ldots, T$. Examples of these type of matrices abound in many areas of science: (i) $X_{it}$  may correspond to the displacements of grain $i=1,\dots,N$ at time $t$ in granular materials; (ii) $X_{it}\in\{0,1\}$  can describe either the neuron activity of neuron $i$ at time $t$ or market activity of stock $i$ at time $t$\footnote{In both cases the time scale depends on the experimental measurements, the frequency of data gathering, etc.}; (iii) $X_{it}$ could stand for the return of stock $i$ at time $t$; (iv) $X_{i\gamma}$  could be the rating of  book $i=1,\ldots,N$ by reader $\gamma$.  Regardless of the research area, let us assume that  the matrix $X$ is generated by some unknown underlying process. This implies that there is an actual correlation matrix $C$ which could be calculated if we were to know the actual intricacies of the underlying world. Let us denote  as $\bracket{\cdots}_{t}$ the theoretical average of such process.  The theoretical $N\times N$ correlation matrix $C$ with entries $C_{ij}=\bracket{x_{i}x_{j}}_t$ for $i,j=1,\ldots,N$, contains information of how the system reacts to perturbations. One way to have access to the system's behaviour is to construct the $N\times N$ empirical correlation matrix $E$ with entries $E_{ij}=\frac{1}{T}\sum_{t=1}^T X_{it}X_{jt}=\frac{1}{T}(X X^T)_{ij}$ and  to do a  Principal Component Analysis\cite{Majumdar2012} that yields estimates for the underlying process. To refine these estimates  one can then exploit universal properties of Random Matrix Theory (RMT) of the Wishart-Laguerre ensemble \cite{Wishart1928},  to obtain, for instance, rotationally invariant estimates \cite{Bouchaud2009,Bun2015}.
 As RMT turns out to be very useful in the analysis of real data, there has been an increased attention to directly  study the various statistical properties related to the Wishart ensemble -as well as other ensembles of random matrices- like, for instance, large deviations of extreme value statistics using the Coulomb fluid picture\cite{Dean2006,Dean2008,Marino2014,Majumdar2012,Vivo2007,Vivo2007b,Vivo2008,Majumdar2009,Majumdar2009b, Fachi2008, Katzav2010, Nadal2010,Nadal2011,Fyodorov2012,Majumdar2013,Majumdar2014, Perez2014, Perez2014b,Perez2014c,Ramli2012,Majumdar2011b,Nadal2009}.\\
The main goal of the present work is to complement the work done in  \cite{Katzav2010,Vivo2007,Majumdar2009} in the Wishart ensemble by using the method presented in \cite{Perez2014,Perez2014b} which is able to capture in a single rate function, and within the Coulomb fluid picture, the statistics of the extreme and bulk eigenvalues, including both the left and right rate functions of the smallest and largest eigenvalues.\\
This paper is organised as follows: in \Sref{sec:Definitions} we introduce the Wishart ensemble and its corresponding joint  probability density function of eigenvalues, from which we want to extract the statistics of the shifted index number and, as explained in the section, the statistics of the $k$-th eigenvalue. In \Sref{sec:Coulomb}, we use the Coulomb fluid picture, originally introduced by Dyson, to derive the statistics of the shifted index number. This entails to applying the saddle-point method and solving the corresponding saddle-point equations  using the Hilbert-Stieltjes transform. The analysis of the latter yields deformations of the spectral density. This deformed Mar\v cenko-Pastur law can then be used to obtain an exact expression of the probability density  of the shifted index number. In \Sref{sec:MonteCarlo}, the analytical results are contrasted with extensive Monte Carlo simulations using Metropolis algorithm directly on the Coulomb fluid. Finally, we conclude with some remarks and possible research lines in \Sref{sec:Conclusions}.

\section{Definitions}
\label{sec:Definitions}
Let us start with some definitions. Given a random variable $X$ taking values $x$ on a given set $x\in\Omega$, we denote as $F_X(x)=$Prob$[X\leq x]$ its cumulative density function (CDF) and $\overline{F}_X(x)=1-F_{X}(x)$ its tail CDF. Similarly, we denote the probability density function (PDF) as $f_{X}(x)=$Prob$[X=x]$.\\
We are interested in studying certain statistical properties related to the joint Probability Density Function (jPDF) of eigenvalues $\bm{y}=(y_1,\ldots,y_N)$ of the Wishart ensemble. This, recall, is defined by starting with rectangular Gaussian matrices $M\times N$, denoted as $\bm{X}$, and forming the so-called Wishart matrices as the $N\times N$ matrices $\bm{W}=\bm{X}^\dag\bm{X}$.  The jPDF of eigenvalues for the latter is known to be:
\beeqn{
P(\bm{y})=\frac{1}{A_0}e^{-\frac{\beta}{2}\sum_{i=1}^{N}y_i}\prod_{i<j}|y_i-y_j|^{\beta}\prod_{i=1}^N y_i^{p-1}\,,
}
where $p=\frac{\beta}{2}(1+M-N)$, $\beta$ is Dyson's index, $A_{0}$ is a normalising factor for $P(\bm{y})$ such that $0<y_i<\infty$ for $i=1,\ldots,N$.  \\
In particular we focus on what we call the shifted index number (SIN) as the random variable $\mathcal{N}_x=\sum_{i=1}^N\Theta(x-y_i)$; this is the number of eigenvalues to the left of $x$ and therefore $\mathcal{N}_x$ can take values on the set $n_{x}\in\{0,\ldots,N\}$. As we will see below the SIN contains more information than the index number previously studied in \cite{Majumdar2009b,Majumdar2011b}. Being this a random variable, a simple expression for its PDF, viz.
\beeqn{
f_{\mathcal{N}_x}(n_x)&=\int_{0}^{\infty}d\bm{y}P(\bm{y})\delta\left(n_x-\sum_{i=1}^N\Theta(x-y_i)\right)\,,
}
would be most welcome. Notice that, even thought the SIN takes discrete values, it is mathematically harmless to consider Dirac deltas instead of Kronecker deltas, at least in the thermodynamic limit. Its tail CDF is obviously given by\footnote{The upper limit in this integral  has been put to $\infty$ while, in principle, it should be $N$. This is not a problem if we understand that either the PDF $f_{\mathcal{N}_x}$ has its support and/or we are already anticipating the thermodynamic limit.}
\beeqn{
\overline{F}_{\mathcal{N}_x}(n_x)=\int_{n_x }^{\infty} d y f_{\mathcal{N}_x}(y)\,.
} 
Next, as noticed in  \cite{Perez2014,Perez2014b}, we do a trivial observation: the probability that the $k$-th eigenvalue $y_k$ is smaller than $x$ is precisely the probability that at least $\mathcal{N}_x$ is greater than $k$, that is
\beeqn{
 \overline{F}_{\mathcal{N}_x}(k)=F_{y_k}(x)\,.
}
Alternatively, we have that $F_{\mathcal{N}_x}(k)=\overline{F}_{y_k}(x)$. Thus, by studying $F_{\mathcal{N}_x}$ (or equivalently $f_{\mathcal{N}_x}$) we have access not only to the statisical properties of SIN, but also to the $k$-th eigenvalue. As pointed out in the work done \cite{Perez2014,Perez2014b} on the Gaussian ensemble, we will see here that we recover the full statistical properties (that is, for large $N$, we obtain the left and right large deviation functions) for both bulk and extreme eigenvalues.

\section{Methods: the Coulomb Fluid approach}
\label{sec:Coulomb}
We use the Coulomb fluid method to derive an expression for $f_{\mathcal{N}_x}(n_x)$ for large $N$ and $M$, while keeping their ratio fixed. We start by rewriting the jPDF of eigenvalues as $P(\bm{y})=(1/A_0)e^{-\frac{\beta}{2} F(\bm{y})}$ with 
\beeqn{
F(\bm{y})=\sum_{i=1}^Ny_i-u\sum_{i=1}^N\ln y_i -\sum_{i\neq j}\ln|y_i-y_j|\,,
}
with $u=\frac{2}{\beta}(p-1)$. Next and with a modest amount of foresight, we introduce rescaled eigenvalues $\lambda_{i}=y_i/N$ to write $F(\bm{y})$ as follows:
\beeqn{
G(\bm{\lambda})\equiv F(N\bm{\lambda})&=N\sum_{i=1}^N\lambda_i-u\sum_{i=1}^N\ln \lambda_i -\sum_{i\neq j}\ln|\lambda_i-\lambda_j|\\
&-uN\ln N-N(N-1)\ln N\,.
}
To go to a continuous theory  we introduce the density of eigenvalues $\rho(\lambda;\bm{\lambda})=(1/N)\sum_{i=1}^N\delta(\lambda-\lambda_i)$ to write $G(\bm{\lambda})=N^2S[\rho(\lambda;\bm{\lambda})]$ with the following expression for large $N$
\beeqn{
\hspace{-2cm}S[\rho(\lambda;\bm{\lambda})]&=\int d\lambda  \rho(\lambda;\bm{\lambda}) (\lambda -\alpha\ln(\lambda)) -\int\int d\lambda d\lambda' \rho(\lambda;\bm{\lambda})\rho(\lambda';\bm{\lambda})\ln|\lambda-\lambda'|\,,
}
where we have defined $\alpha= u/N$, which, for $N$ and $M$ large, gives $\alpha=(M-N)/N=(1-d)/d$ with $d=N/M$. This allows us to write $P(\bm{y})=(1/A_0)e^{-\frac{\beta}{2}N^2 S[\rho(\lambda;\bm{\lambda})]}$.\\
To derive an expression of the PDF of the SIN, we introduce intensive variables $c=n_x/N$ and denote $\varrho(c)= (1/N)f_{\mathcal{N}_x}(n_x/N)$ to write
\beeqn{
\varrho(c)&=\frac{1}{Z_0}\int D[\rho]e^{-\frac{\beta}{2}N^2 S[\rho(\lambda)]}\delta\left(c-\int d\lambda \rho(\lambda)\Theta(x-\lambda)\right)\\
&\times\int_0^\infty d\bm{\lambda}\delta_{(F)}\left(\rho(\lambda)-\frac{1}{N}\sum_{i=1}^N\delta(\lambda-\lambda_i)\right)\,.
}
The latter term gives rise to an entropic contribution. After evaluating this part one ends up with the following formula for $\varrho(c)$:
\beeq{
\varrho(c)&=\frac{1}{Z_0}\int D[\rho,B_1,B_2]e^{-\frac{\beta}{2}N^2\mathcal{A}[\rho,B_1,B_2]}\,,
\label{erho}
}
with 
\beeq{
\hspace{-1cm}\mathcal{A}[\rho,B_1,B_2]&= \int d\lambda  \rho(\lambda) (\lambda -\alpha\ln(\lambda)) -\int\int d\lambda d\lambda' \rho(\lambda)\rho(\lambda')\ln|\lambda-\lambda'|\nonumber\\
\hspace{-1cm}&+B_1\left(\int d\lambda \rho(\lambda)\Theta(x-\lambda)-c\right)+B_2\left(\int d\lambda \rho(\lambda)-1\right)\,,
\label{erho2}
}
where $ D[\rho,B_1,B_2]$ stands for path integral for  the density $\rho(\lambda)$ and standard integration over the variables $B_1$ and $B_2$. In this derivation various constants have been absorbed into the normalising constant $Z_0$. The rest to be done is to get an expression of $\varrho(c)$ for large $N$ by analysing \eref{erho} and \eref{erho2}  using the saddle-point method. Looking at eq. \eref{erho2} this entails to obtaining the equilibrium distribution of the charge density of a two-dimensional Coulomb fluid restricted on a one-dimensional line with a fraction charge $c$ constrained to be to the left of a barrier at position $x$. This yields fairly intuitively  to deformations of the Mar\v cenko-Pastur law.\\
Being more precise, in the limit of $N$ large we  write
\beeq{
\varrho(c)=e^{-\beta N^2\Psi(c,x)}\,,\quad \Psi(c,x)=\frac{1}{2}\left(\mathcal{A}_0(c,x)-\mathcal{A}_{{\tt MP}}\right)\,,
}
where $\Psi$ is the so-called rate function. Here we have used that at the saddle point the numerator and denominator appearing in \eref{erho} can be written as $e^{-\frac{\beta}{2}N^2\mathcal{A}_0(c,x)}$ and $Z_{0}=e^{-\frac{\beta}{2}N^2\mathcal{A}_{{\tt MP}}}$, respectively.  $\mathcal{A}_0(c,x)$ corresponds to the value of the action $\mathcal{A}[\rho,B_1,B_2]$ evaluated at the saddle-point and, therefore, it will depend on the parameters $c$ and $x$ (as well as on $\alpha$). Similarly, $\mathcal{A}_{{\tt MP}}$ corresponds to the evaluation of the  action associated with $Z_0$, which is the equilibrium state of the Coulomb fluid yielding the Mar\v cenko-Pastur law. All in all, we see that $\varrho(c)$ is simply related to equilibrium  distributions of the Coulomb fluid in presence of the constraint $\int d\lambda \rho(\lambda)\Theta(x-\lambda)=c$. If we are able to find this deformed Mar\v cenko-Pastur density and evaluate its corresponding free energy, then we will be able to  derive  an expression for the rate function. From $\varrho(c)=e^{-\beta N^2\Psi(c,x)}$ its tail CDF reads
\beeqn{
\overline{F}_{\mathcal{C}_x}(c)={\rm Prob}[\mathcal{C}_x\geq c]=\int_{c}^1 dc'\varrho(c')\,.
}
As we will see below  the PDF $\varrho(c)$ is peaked at  a particular value of $c=c_\star(x)$. Then we have that:
\beeqn{
\overline{F}_{\mathcal{C}_x}(c)&=e^{-\beta N^2\Psi(c,x)}\,,\quad c>c_{\star}(x)\\
F_{\mathcal{C}_x}(c)&=e^{-\beta N^2\Psi(c,x)}\,,\quad c<c_{\star}(x)\,.
}
Once that the CDF of the SIN is found, we automatically obtain the corresponding CDF for the $k$-th eigenvalue, \textit{viz.}
\beeqn{
F_{\lambda_k}(x)&=e^{-\beta N^2\Psi(k/N,x)}\,,\quad x<x_{\star}(k/N)\\
\overline{F}_{\lambda_k}(x)&=e^{-\beta N^2\Psi(k/N,x)}\,,\quad x>x_{\star}(k/N)\,.
\label{eq:LDk}
}
Thus, the rate function $\Psi(c,x)$ has a two-fold meaning: as a function of $c$ (for $x$ fixed) gives information about the large deviations of the SIN; as a function of $x$ (for $c=k/N$ fixed) gives the large deviations of the $k$-th eigenvalue. The latter case is even more surprising when, as we will see later, we find that for extreme eigenvalues, that is for $k=1$ (smallest) or $k=N$ (largest), one is able to derive their left and right large deviation functions within the Coulomb fluid picture.

\subsection{The saddle-point equations}
Doing a variation of the action, viz. 
\beeq{
\frac{\delta \mathcal{A}[\rho,B_1,B_2]}{\delta \rho (\lambda)}=\frac{\partial \mathcal{A}[\rho,B_1,B_2]}{\partial B_1}=\frac{\partial \mathcal{A}[\rho,B_1,B_2]}{\partial B_2}=0\,,
}
yields the following saddle-point  equations
\beeq{
&(\lambda-\alpha\ln(\lambda))+B_1\Theta(x-\lambda)+B_2=2\int d\lambda'\rho(\lambda')\ln|\lambda-\lambda'|\label{eq:spe1}\,,\\
&c=\int d\lambda \rho(\lambda)\Theta(x-\lambda)\,,\quad \quad  1=\int d\lambda \rho(\lambda)\,.\label{eq:spe2}
}
Henceforth one typically does the following manipulations: first of all we perform the derivative with respect to $\lambda$ in \eref{eq:spe1}; the resulting equation is multiplied by  $\rho(\lambda)/(z-\lambda)$ and integrated over $\lambda$; next one introduces the Hilbert-Stieltjes transform $S(z)=\int d\lambda \frac{\rho(\lambda)}{z-\lambda}$ of the density which, after some manipulations yields the following second order polynomial equation for $S(z)$
\beeq{
S^2(z)=S(z)-\frac{\alpha}{ z}S(z)+\frac{\gamma}{ z}+\frac{\omega}{z-x}\,,
\label{eq:2op}
}
where $\gamma$ and $\omega$ are two parameters chosen appropriately to eliminate unnecessary constants.  Solving \Eref{eq:2op} and imposing that $S(z)\sim 1/z$ for $z\to \infty$, relates the two parameters as $\gamma=-1-\omega$. After some final manipulations the resolvent takes the following form
\beeq{
S_{\pm}(z)\equiv S_{\pm}(z;x,\omega;\alpha)=\frac{1}{2z}\left[\left(z-\alpha\right)\pm\sqrt{\frac{P_3(z)}{z-x}}\right]\,,
\label{eq:2}
}
with the cubic polynomial
\beeq{
P_{3}(z)=(z-x)(z-b_{+})(z-b_{-})+4xz\omega\,,
}
and $b_{\pm}\equiv b_{\pm}(\alpha)=(1\pm\sqrt{1+\alpha})^2$ being the natural upper and lower limit supports of the Mar\v cenko-Pastur law. Note  that when $\omega=0$ we recover the resolvent associated to the Mar\v cenko-Pastur law $\rho_{{\tt MP}}(\lambda)=\frac{\sqrt{(b_{+}(\alpha)-\lambda)(\lambda-b_{-}(\alpha))}}{2\pi \lambda}I_{\lambda\in[b_{-}(\alpha),b_{+}(\alpha)]}$, viz
\beeq{
S_{\pm}^{{\tt MP}}(z)=\frac{1}{2z}\left[\left(z-\alpha\right)\pm\sqrt{(z-b_{+})(z-b_{-})}\right]\,,
}
and therefore we conclude that the parameter $\omega$ controls the deformation of the Mar\v cenko-Pastur law due to the barrier at the position $x$ and the fraction of eigenvalues $c$ to the left of it. Looking at the resolvent \eref{eq:2}, everything boils down to analysing the roots of the polynomial $P_3(z)$ as a function of the parameters of the problem.
\subsection{Analysis of the Roots of $P_3(z)$}
\begin{figure*}[t]
\includegraphics[width=5cm,height=5cm]{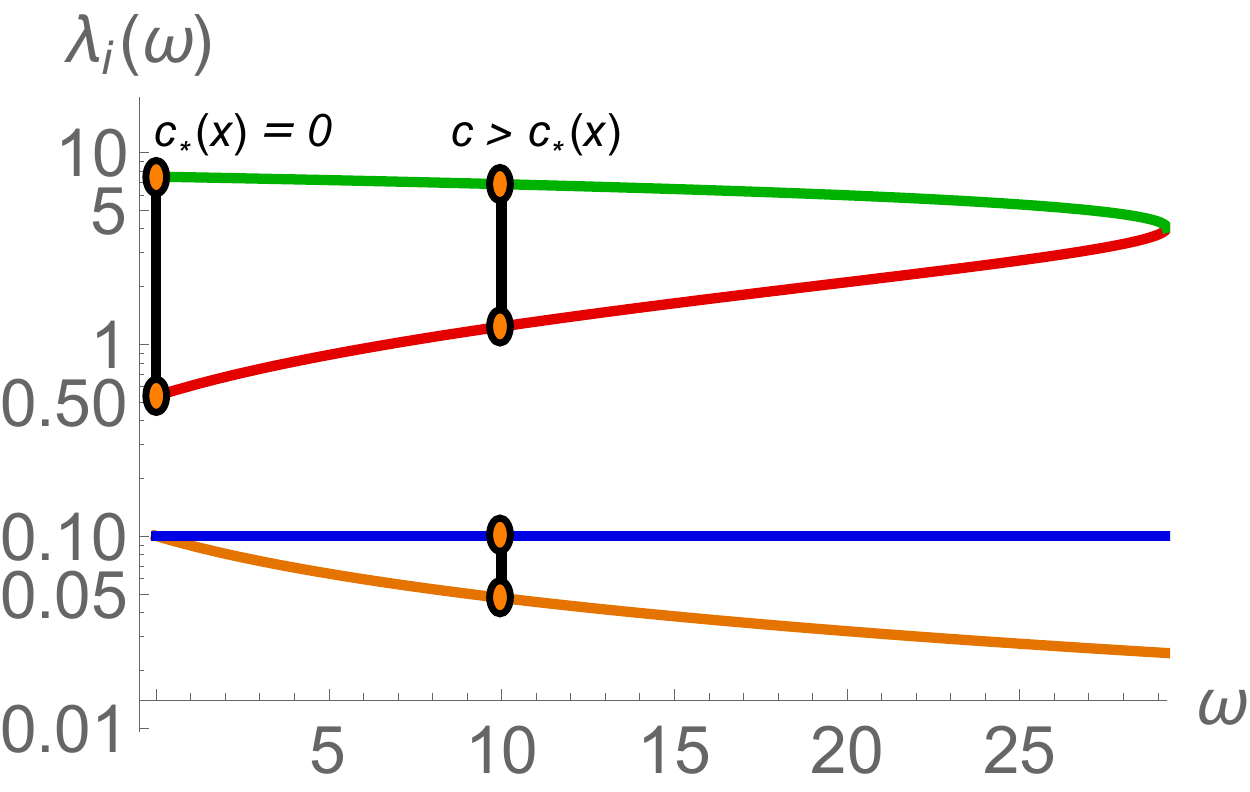}\quad
\includegraphics[width=5cm,height=5cm]{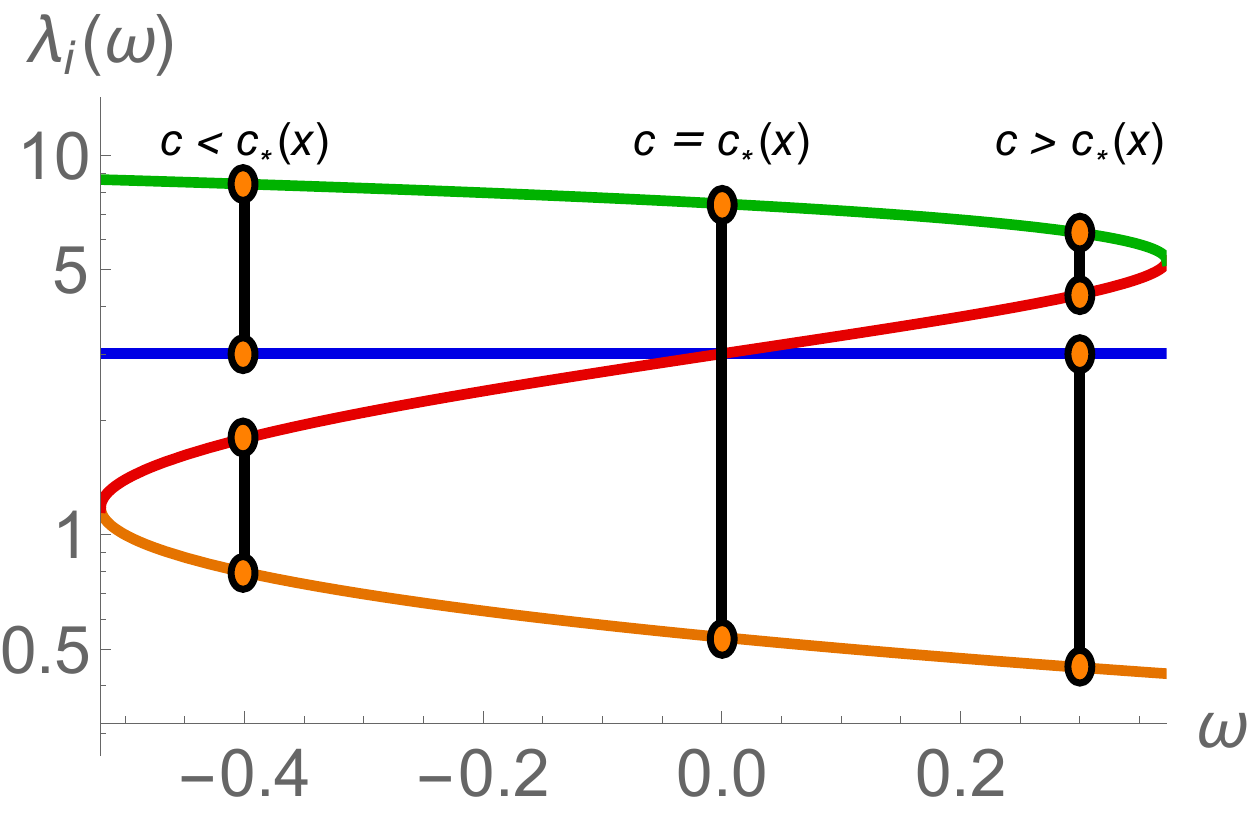}\quad
\includegraphics[width=5cm,height=5cm]{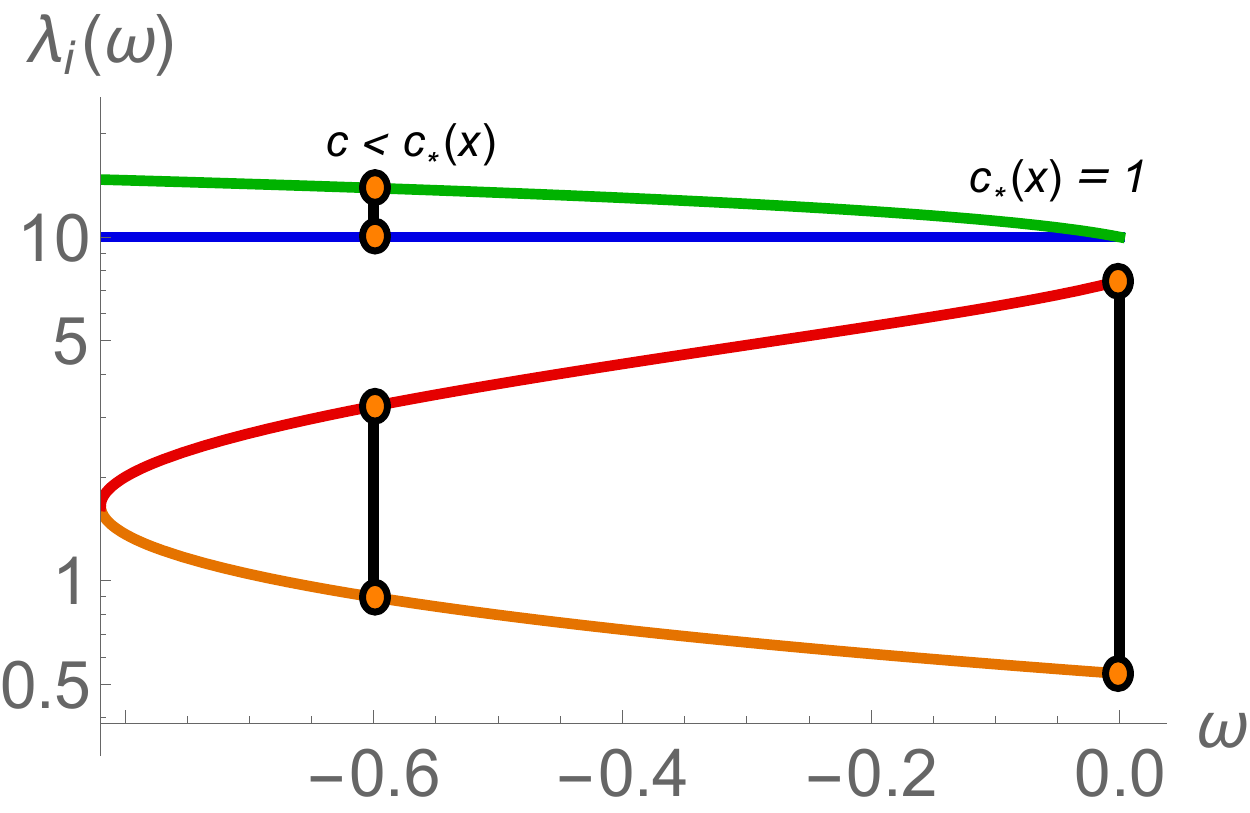}
\caption{Plot of the roots  $0\leq \lambda_{-}\leq \lambda_0\leq \lambda_{+}$ and the position of the barrier (solid blue line) as a function of $\omega$ while fixing the value of  $\alpha=2$ and $x=1/10,3,10$ (from left to right).}
\label{fig1}
\end{figure*}

\begin{figure*}[t]
\includegraphics[width=7cm,height=5cm]{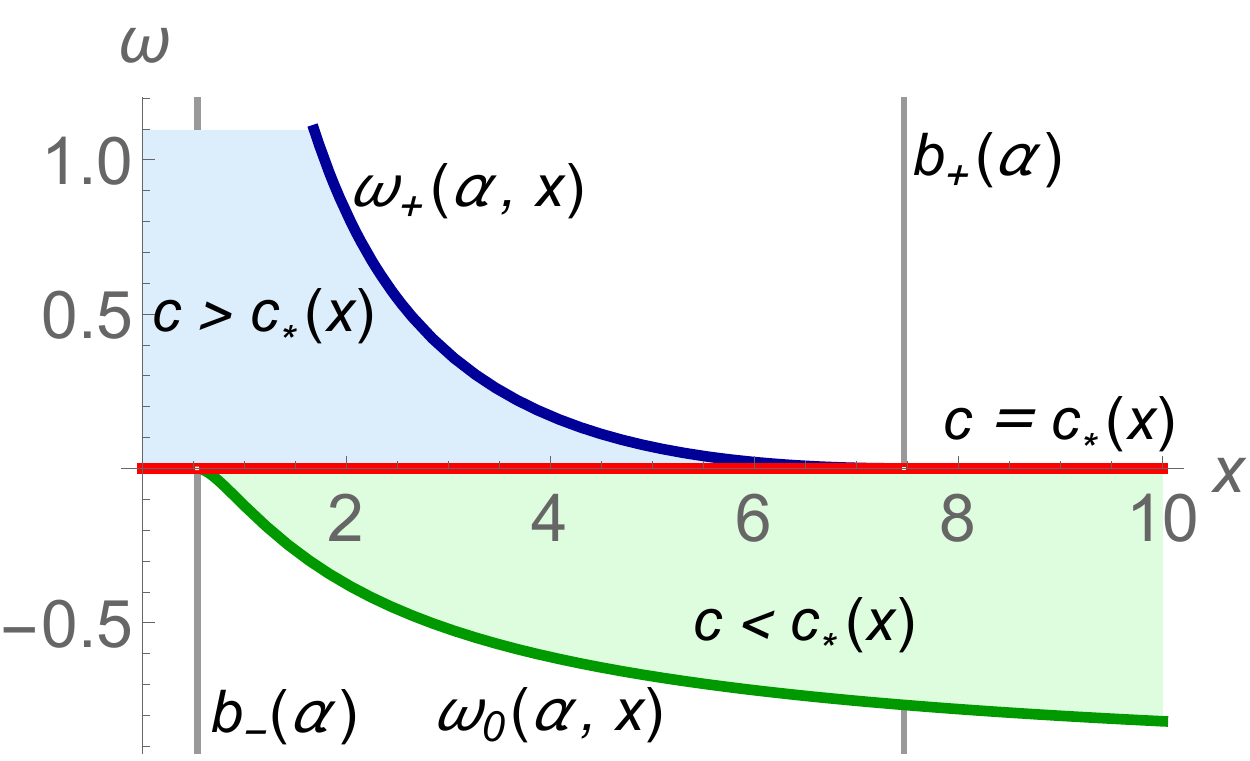}\quad\quad\quad\includegraphics[width=7cm,height=5cm]{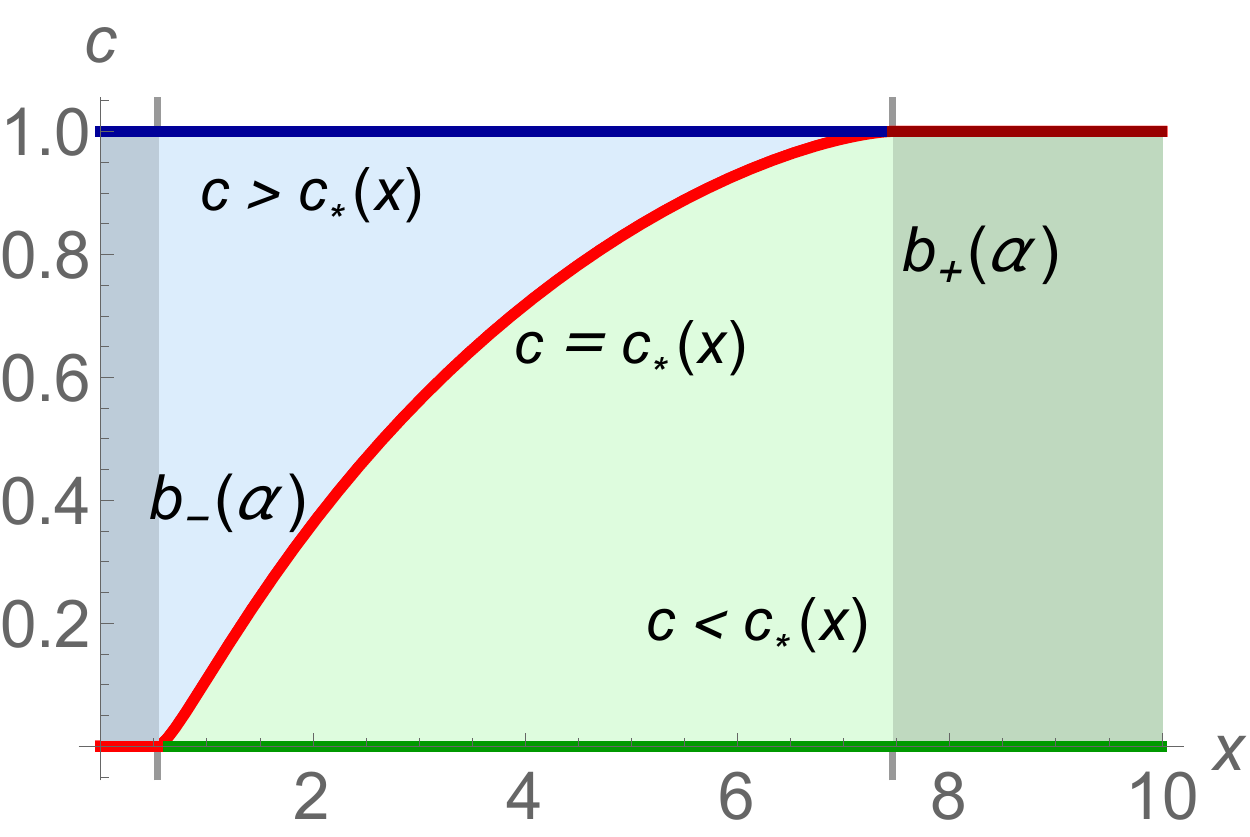}
\caption{}
\label{xo}
\end{figure*}

The discriminant $\Delta$ of the cubic equation $P_3(z)=0$ can be written as follows
\beeq{
\Delta=-256x^3(\omega-\omega_{0}(\alpha,x))(\omega-\omega_{+}(\alpha,x))(\omega-\omega_{-}(\alpha,x))\,,
}
where $\omega_i(\alpha,x)$ are the roots of the $\Delta$ given by\footnote{The roots are always reals and in principle there is no need to take the real part as it appears in their expressions. However, this is convenient if one wants to keep them ordered.}
\beeq{
\hspace{-1cm}\omega_{+}(\alpha,x)&=\frac{1}{48x}\Re\Bigg[\kappa(\alpha,x)-\frac{-1+i \sqrt{3}}{2} \chi(\alpha,x)-\frac{-1-i\sqrt{3}}{2}\frac{\Xi(\alpha,x)}{\chi(\alpha,x)}\Bigg]\,,\\
\hspace{-1cm}\omega_{-}(\alpha,x)&=\frac{1}{48x}\Re\Bigg[\kappa(\alpha,x)- \chi(\alpha,x)-\frac{\Xi(\alpha,x)}{\chi(\alpha,x)}\Bigg]\,,\\
\hspace{-1cm}\omega_0(\alpha,x)&=\frac{1}{48x}\Re\Bigg[\kappa(\alpha,x)-\frac{-1-i \sqrt{3}}{2}\chi(\alpha,x) -\frac{-1+i\sqrt{3}}{2}\frac{\Xi(\alpha,x)}{\chi(\alpha,x)}\Bigg]\,,
}
with  definitions
\beeq{
\hspace{-1cm}\Gamma(\alpha,x)&=-64 \alpha ^6+96 \alpha ^5 (43 x-8)+24 \alpha ^4 (x (503 x+1000)-160)\nonumber\\
\hspace{-1cm}&+40 \alpha ^3 (77 x-16) (x+4)^2+240 \alpha ^2 (2 x-1) (x+4)^3\nonumber\\
\hspace{-1cm}&-12 \alpha  (x+4)^5-(x+4)^6\,,\\
\hspace{-1cm}\Omega(\alpha,x)&=24 \sqrt{3}\sqrt{\alpha ^2 (-x) \left(8 \alpha ^3+3 \alpha ^2 (16-5 x)+6 \alpha  (x+4)^2+(x+4)^3\right)^3}\,,\\
\hspace{-1cm}\Xi(\alpha,x)&=16 \alpha ^4+16 \alpha ^3 (29 x+8)+48 \alpha ^2 (5 x+2) (x+4)\nonumber\\
\hspace{-1cm}&+8 \alpha  (x+4)^3+(x+4)^4\,,\\
\hspace{-1cm}\chi(\alpha,x)&=\sqrt[3]{\Gamma+\Omega}\,,\quad \kappa(\alpha,x)=-8 ((\alpha -2) \alpha -2)+x^2-20 (\alpha +2) x\,.
}
These roots  are ordered as $\omega_{+}\geq0\geq\omega_{0}\geq \omega_{-}$. This allows us to express the roots of $P_3(\lambda)$ as follows
\beeq{
\hspace{-1cm}\lambda_{-}(\alpha,x,\omega)&=- \frac{1}{3}\left(-4-x-2\alpha + C +\frac{\Delta_0}{C}\right)\,,\\
\hspace{-1cm}\lambda_{0}(\alpha,x,\omega)&=- \frac{1}{3}\left(-4-x-2\alpha + {-1 - i\sqrt{3} \over 2}C +{-1 + i\sqrt{3} \over 2} \frac{\Delta_0}{C}\right)\,,\\
\hspace{-1cm}\lambda_{+}(\alpha,x,\omega)&=- \frac{1}{3}\left(-4-x-2\alpha + {-1 + i\sqrt{3} \over 2}C +{-1 - i\sqrt{3} \over 2} \frac{\Delta_0}{C}\right)\,,
}
with definitions
\beeq{
\hspace{-2cm}C(\alpha,x,\omega) &= \sqrt[3]{\Delta_1 + 24\sqrt{3 x^3(\omega-\omega_{0}(\alpha,x))(\omega-\omega_{+}(\alpha,x))(\omega-\omega_{-}(\alpha,x))}}\,,\\
\hspace{-2cm}\Delta_0(\alpha,x,\omega)&=16+x^2+\alpha(16+\alpha)-2x (2+\alpha+6\omega)\,,\\
\hspace{-2cm}\Delta_1(\alpha,x,\omega)&=-\frac{27}{2} x\alpha^2-(4+x+2\alpha)^3+\frac{9}{2}(4+x+2\alpha)(\alpha^2+2x(2+\alpha+2\omega))\,.
}
 As written, and as one can observe by plotting the roots of $P_3(\lambda)$  as a function of $\omega$ (see  \Fref{fig1}), they appear ordered as $0\leq \lambda_{-}\leq \lambda_0\leq \lambda_{+}$.  From here we see that two scenarios emerge: the first one corresponds when the position $x$ of the barrier is within the natural support of the Mar\v cenko-Pastur law, that is $x\in[b_{-}(\alpha),b_{+}(\alpha)]$ (this corresponds to the middle panel in  \Fref{fig1}). Then, depending on the value of $c$ compared to the natural fraction $c_{\star}(x)$ of eigenvalues of the  Mar\v cenko-Pastur law  to the left of $x$ we may have a doubly supported density (for $c\neq c_{\star}(x)$) or the  Mar\v cenko-Pastur law which corresponds when the constraint is ineffective, that is $c=c_\star(x)$. In this case one observes that $\omega\in[\omega_0(\alpha,x),\omega_{+}(\alpha,x)]$ and the value of $\omega$ is actually controlling the fraction $c$ of eigenvalues to the left of $x$, going from $c=0$ for $\omega=\omega_0(\alpha,x)$ to $c=1$ for $\omega=\omega_{+}(\alpha,x)$. The actual expression  relating $\omega$ and $c$ will be derived later; the second scenario corresponds when the position of the barrier is outside the natural support $x\not\in[b_{-}(\alpha),b_{+}(\alpha)]$. In this case we have that either $\omega\in[\omega_0(\alpha),0]$ or $\omega\in[0,\omega_{+}(\alpha)]$, respectively (and it corresponds to the left and right panels of  \Fref{fig1}).
Actually both scenarios can be brought together by redefining the roots of the discriminant $\Delta$ as follows:
\beeq{
\hspace{-1.5cm}\omega_0(\alpha,x)&=\left\{
\begin{array}{ll}
0&x\leq b_{-}(\alpha)\\
\frac{1}{48x}\Bigg[\kappa(\alpha,x)-\frac{-1-i \sqrt{3}}{2}\chi(\alpha,x) -\frac{-1+i\sqrt{3}}{2}\frac{\Xi(\alpha,x)}{\chi(\alpha,x)}\Bigg]& x\geq b_{-}(\alpha)
\end{array}
\right.\\
\hspace{-1.5cm}\omega_+(\alpha,x)&=\left\{
\begin{array}{ll}
\frac{1}{48x}\Bigg[\kappa(\alpha,x)-\frac{-1+i \sqrt{3}}{2} \chi(\alpha,x)-\frac{-1-i\sqrt{3}}{2}\frac{\Xi(\alpha,x)}{\chi(\alpha,x)}\Bigg]&x\leq b_{+}(\alpha)\\
0&x\geq b_{+}(\alpha)
\end{array}\,.
\right.
}
All in all, a  summary can be found in \Fref{xo}, where one can see the restricted regions in the $(x,\omega)$- and the $(x,c)$-planes.  Here the red curve corresponds to the expression of $c_{\star}(x)=\int_{-\infty}^x d\lambda\rho_{{\tt MP}}(\lambda)$ whose exact expression is  given by
\beeq{
\hspace{-2cm}c_{\star}(x)&=\left\{
\begin{array}{ll}
0&x<b_{-}\\
\frac{\sqrt{ (b_{+}-x) (x-b_{-})}+ 2(2+\alpha) \sin ^{-1}\left(\sqrt{\frac{x-b_{-}}{4\sqrt{1+\alpha}}}\right)-2 \alpha \tan ^{-1}\left(\sqrt{\frac{b_{+} (x-b_{-})}{b_{-} (b_{+}-x)}}\right)}{2 \pi  }& x\in[b_{-},b_{+}]\\
1&x>b_{+}
\end{array}
\right.\,,
}
where from now on we will consider that $M\geq N$ which implies $0\leq d\leq 1$ or $\alpha\in[0,\infty)$.

\subsection{The deformed Mar\v cenko-Pastur law}
With the help of the previous analysis we can finally write an exact expression for the deformed MP law

\begin{figure}[t]
\begin{center}
\includegraphics[width=10cm,height=7cm]{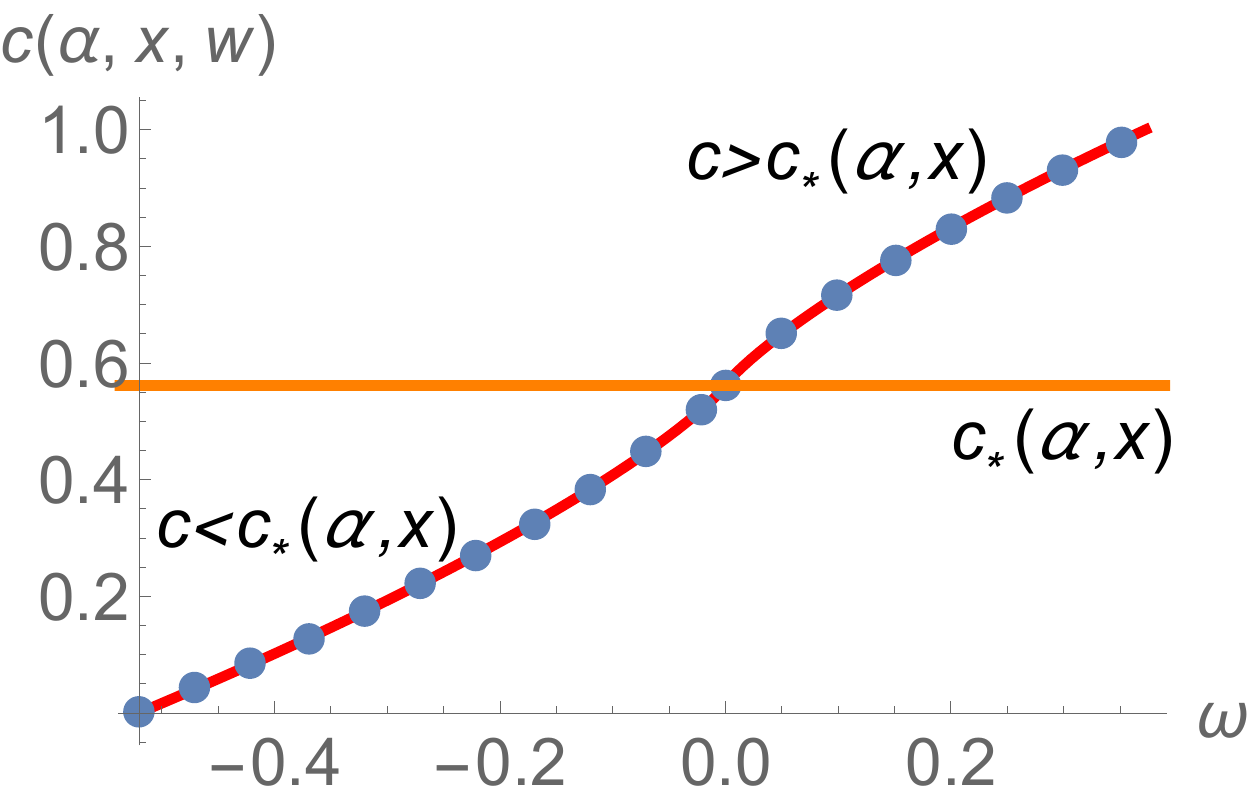}
\caption{Plot of $c$ vs $\omega$ for $x=3$ and $\alpha=2$. Here we compare the theoretical expressions given in the text (solid red lines) with the numerical evaluation of the integral defintion of $c$ in terms of the constrained spectral density (solid blue circles)}
\label{fig:cvsw}
\end{center}
\end{figure}

\begin{figure}[t]
\begin{center}
\includegraphics[width=4cm,height=4.5cm]{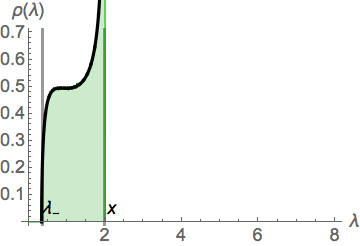}~
\includegraphics[width=4cm,height=4.5cm]{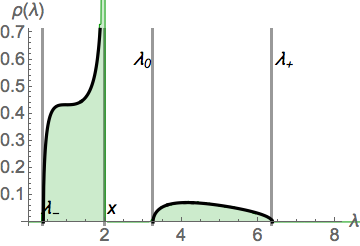}~
\includegraphics[width=4cm,height=4.5cm]{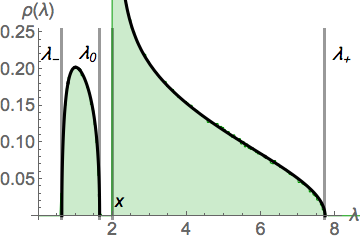}~
\includegraphics[width=4cm,height=4.5cm]{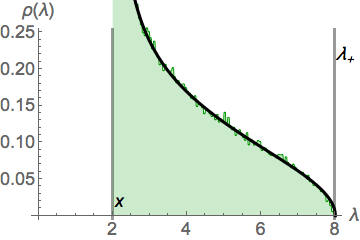}
\caption{Deformed MP laws for $x=2$ and $\alpha=2$  corresponding to the value $c_{\star}(\alpha,x)=0.362418$. The green filled part corresponds to MC simulations with $N=600$  and with $10^5$ MC steps for thermalising and $10^5$ MC steps for gathering all eigenvalues every 100 MC steps. The eigenvalues have been distributed to each site of the barrier to achieve the desired  fraction $c$ that, from left to right, corresponds to the following values $c=1,5/6,1/6$ and $1$.}
\label{fig:dmpls}
\end{center}
\end{figure}

\beeq{
\hspace{-2.8cm}\rho(\lambda)&=\frac{1}{2\pi\lambda}\left\{\begin{array}{ll}
\sqrt{\frac{(\lambda-\lambda_+)(\lambda-\lambda_0)(\lambda-\lambda_-)}{x-\lambda}}I_{\lambda\in[\lambda_{-},x]}+\sqrt{\frac{(\lambda_+-\lambda)(\lambda-\lambda_0)(\lambda-\lambda_-)}{\lambda-x}}I_{\lambda\in[\lambda_{0},\lambda_{+}]}& c>c_{\star}(x)\\
\sqrt{\frac{(\lambda-\lambda_+)(\lambda_0-\lambda)(\lambda-\lambda_-)}{\lambda-x}}I_{\lambda\in[\lambda_{-},\lambda_0]}+\sqrt{\frac{(\lambda_+-\lambda)(\lambda-\lambda_0)(\lambda-\lambda_-)}{\lambda-x}}I_{\lambda\in[x,\lambda_{+}]}& c<c_{\star}(x)
\end{array}\,.
\label{eq:dmp}
\right.
}
This expression is simplified significantly when all eigenvalues are either to the left or to the right of the barrier. From the point of view of the roots of $P_3(\lambda)$ this algebraically corresponds to having $\Delta=0$ and   $\Delta_0 \neq 0$, which implies in turn to have a double root and a single root, viz.
\beeq{
\hspace{-2cm}\lambda_{1}(\alpha,x,\omega)=\frac{\alpha ^2 (\alpha +2)+x^2 (\alpha +2 \omega +2)+x \left(-2 \alpha ^2+4 \alpha  (\omega +2)+8 (\omega +1)\right)}{\alpha  (\alpha +16)+x^2-2 x (\alpha +6 \omega +2)+16}\,,\\
\hspace{-2cm}\lambda_{2}(\alpha,x,\omega)=\frac{-4 (2 \alpha +x+4) \left(\alpha ^2+2 x (\alpha +2 \omega +2)\right)+9 \alpha ^2 x+(2 \alpha +x+4)^3}{\alpha  (\alpha +16)+x^2-2 x (\alpha +6 \omega +2)+16}\,,
}
respectively. Here $\omega$ must be chosen so that either $\omega=\omega_0$ (corresponding to all eigenvalues to the right) or $\omega=\omega_+$ (corresponding to all eigenvalues to the left). Notice that graphically the root that doubles corresponds to the ones attached to the barrier. Thus for $c=1$ we have that $\lambda_{+}=\lambda_{0}$ (so that the upper blob disappears) while for $c=0$ we have that $\lambda_{-}=\lambda_{0}$. All in all, the following equations summarise the results. The spectral density for all eigenvalues to the left or to the right of the barrier $x$ has the following form
\beeq{
\rho_{L}(\lambda)=\left\{
\begin{array}{ll}
\rho_{{\rm MP}}(\lambda)& x\geq b_{+}(\alpha)\\
\frac{1}{2\pi\lambda}\sqrt{\frac{(\lambda-\ell_L(\alpha,x))}{x-\lambda}}\left|\lambda-u_L(\alpha,x)\right|I_{\lambda\in[\ell_L(\alpha,x),x]}& x\leq b_{+}(\alpha)
\end{array}\,,
\right.
}
and
\beeq{
\rho_{R}(\lambda)=\left\{
\begin{array}{ll}
\rho_{{\rm MP}}(\lambda)& x\leq b_{-}(\alpha)\\
\frac{1}{2\pi\lambda}\sqrt{\frac{u_R(\alpha,x)-\lambda}{\lambda-x}}\left|\lambda-\ell_R(\alpha,x)\right|I_{\lambda\in[x,u_R(\alpha,x)]}& x\geq b_{-}(\alpha)
\end{array}
\right.\,,
}
respectively, and it agrees with the results already presented in \cite{Dean2006}. Here we have used the following definitions:
\beeq{
\hspace{-2cm}u_L(\alpha,x)&=\frac{\alpha ^2 (\alpha +2)+x^2 (\alpha +2 \omega_{+} +2)+x \left(-2 \alpha ^2+4 \alpha  (\omega_{+} +2)+8 (\omega_{+} +1)\right)}{\alpha  (\alpha +16)+x^2-2 x (\alpha +6 \omega_{+} +2)+16}\,,\\
\hspace{-2cm}\ell_L(\alpha,x)&= \lambda_{-}(\alpha,x,\omega_{+}(\alpha,x))\,,\\
\hspace{-2cm}\ell_R(\alpha,x)&=\frac{\alpha ^2 (\alpha +2)+x^2 (\alpha +2 \omega_0 +2)+x \left(-2 \alpha ^2+4 \alpha  (\omega_{0} +2)+8 (\omega_{0} +1)\right)}{\alpha  (\alpha +16)+x^2-2 x (\alpha +6 \omega_{0} +2)+16}\,,\\
\hspace{-2cm}u_R(\alpha,x)&= \lambda_{+}(\alpha,x,\omega_{0}(\alpha,x))\,.
}
We are finally left to obtain an expression relating the parameter $\omega$ with the fraction $c$ of eigenvalues to the left of the barrier $x$. This is done simply by integrating over the first blob in the deformed MP laws given by eqs. \eref{eq:dmp}. The corresponding integral can be expressed in terms of elliptic integrals. After a lengthy and tedious derivation one eventually arrives at:
\beeq{
\hspace{-2cm}c(\alpha,x,w)&=\frac{1}{2\pi\sqrt{(a-c) (b-d)}}\Bigg[(a-c) (b-d) E\left(k\right)-(a-d) \Big((a-c) K\left(k\right)\\
\hspace{-2cm}&-(a+b-c+d) \Pi \left(\frac{d-c}{a-c},k\Big)+2 b \Pi \left(\frac{a (d-c)}{d(a-c) },k\right)\right)\Bigg]\,,
}
for $c>c_{\star}(x)$ and with  $a=\lambda_{+}$, $b=\lambda_0$ $c=x$, and $d=\lambda_{-}$. For $c<c_{\star}(x)$ we obtain instead
\beeq{
\hspace{-2cm}c(\alpha,x,w)&=\frac{1}{b \sqrt{(a-c) (b-d)}}\Bigg[(b-c) \Bigg(-(2 a-b) (b-d) K\left(k\right)\nonumber\\
\hspace{-2cm}&+b (a-b+c+d) \Pi \left(\frac{c-d}{b-d},k\right)\nonumber\\
\hspace{-2cm}&-2 a d \Pi \left(\frac{b (c-d)}{c (b-d)},k\right)\Bigg)+b (a-c) (b-d) E\left(k\right)\Bigg]\,,
}
with $a=\lambda_+$, $b=x$, $c=\lambda_0$, and $d=\lambda_-$. In both cases  $k=\sqrt{\frac{(a-b)(c-d)}{(a-c)(b-d)}}$ is the elliptic modulus, while $K(k)$, $E(k)$, and $\Pi(n,k)$ correspond to the complete elliptic integrals of the first second and third kind, respectively. In plot \Fref{fig:cvsw} we show an example of $c$ as a function of $\omega$ and contrast it with the numerical integration of the deformed spectral density.\\
We have now all the necessary expressions to plot the deformed MP laws. This is shown in  \Fref{fig:dmpls} where we plot the exact expressions together with Monte Carlo simulations for decreasing values of $c$ from left to right.
\subsection{Evaluating the action and the rate function at the saddle point}
Using the saddle point eq. \eref{eq:spe1}, the action evaluated at the saddle point takes the following form 
\beeq{
\mathcal{A}_{0}(c,x)=\frac{1}{2}\int d\lambda  \rho(\lambda) (\lambda -\alpha\ln(\lambda)) -\frac{1}{2}B_1 c-\frac{1}{2}B_2\,.
\label{eq:asp}
}
It is possible to obtain integral expressions for the contants $B_1$, $B_2$ and for the first term in formula \eref{eq:asp} in terms of the resolvent. Indeed, the first term of \eref{eq:asp} corresponds to the combination of the first moment of $\rho(\lambda)$ and the average $\bracket{\ln(\lambda)}_{\rho}$. The first moment can be obtained by an expansion of the resolvent $S(z)$ in inverse powers of $z$ , viz. $S(z)=\sum_{\ell=0}^\infty\frac{\mu_\ell}{z^{\ell+1}}$ with $ \mu_\ell=\int d\lambda \rho(\lambda)\lambda^\ell$. In our case we have that $\mu_1=1+\alpha-x\omega$. To derive an formula for $\bracket{\ln(\lambda)}_{\rho}$ we do the following trick:   integrating the resolvent in the interval $z\in[0,\lambda_{-}]$ yields:
\beeq{
\int d\lambda  \rho(\lambda) \ln\left|\lambda\right|=\int d\lambda  \rho(\lambda) \ln\left|\lambda_--\lambda\right|-\int_{0}^{\lambda_{-}} dz S_{+}(z)\,.
}
Moreover, taking $\lambda=\lambda_{-}$ at the saddle-point equation \eref{eq:spe1} and after combining both expressions results into the formula
\beeq{
\int d\lambda  \rho(\lambda) \ln\left(\lambda\right)=\frac{1}{2}\left[B_1+B_2+\lambda_{-}-\alpha\ln(\lambda_{-})\right]-\int_{0}^{\lambda_{-}} dz S_{+}(z)\,.
}
All in all, the first term in \eref{eq:asp} can be expressed as
\beeq{
\hspace{-2cm}\frac{1}{2}\int d\lambda  \rho(\lambda) (\lambda -\alpha\ln(\lambda))&=\frac{1+\alpha-x\omega}{2}\nonumber\\
&-\frac{\alpha}{4}\left[B_1+B_2+\lambda_{-}-\alpha\ln(\lambda_{-})-2\int_{0}^{\lambda_{-}} dz S_{+}(z)\right]\,.
}
A similar analysis can be done to express the constants $B_1$ and $B_2$ in terms of the resolvent. Combining these results the final expression for the action is:
\beeq{
\hspace{-1.5cm}\mathcal{A}_{0}(c,x)&=\frac{1+\alpha-x\omega}{2}-\frac{\alpha}{4}\left(\lambda_{-}-\alpha\ln(\lambda_{-})-2\int_{0}^{\lambda_{-}} dz S_{+}(z)\right)\nonumber\\
\hspace{-1.5cm}&-\frac{1}{2}\left(c+\frac{\alpha}{2}\right)\left(x-\lambda_{0}+\alpha\ln(\lambda_{0}/x)+2\int_{x}^{\lambda_0} dz S_{-}(z)\right)\label{eq:af} \\
\hspace{-1.5cm}&-\frac{1}{2}\left(1+\frac{\alpha}{2}\right)\left(-\lambda_{+}+(2+\alpha)\ln(\lambda_{+})-2\int_{\lambda_{+}}^\infty\left[S_{-}(z)-\frac{1}{z}\right]\right)\nonumber\,,
}
which is valid for  $c>c_{\star}$ and $c<c_{\star}$. The rate function is just $\Psi(c,x)=\frac{1}{2}\left(\mathcal{A}_0(c,x)-\Omega_0\right)$ with
\beeq{
\Omega_0&=\frac{1}{2} \left(\alpha ^2 \ln \left(\frac{\alpha }{\alpha +1}\right)+3 \alpha -(2 \alpha +1) \ln (\alpha +1)+3\right)\,.
}
\begin{figure*}[t]
\begin{center}
\includegraphics[width=7cm,height=7cm]{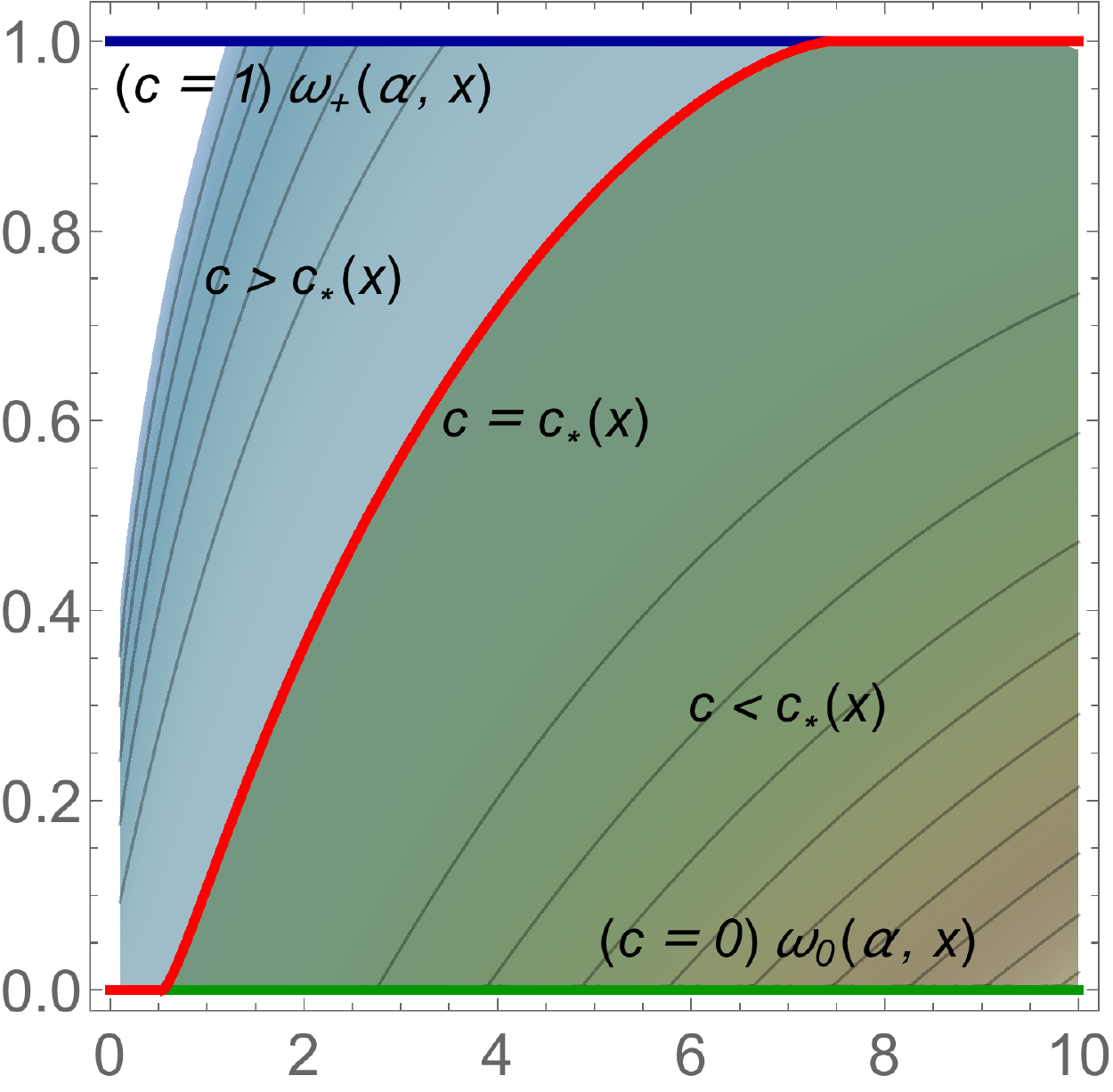}\quad\quad\includegraphics[width=7cm,height=7cm]{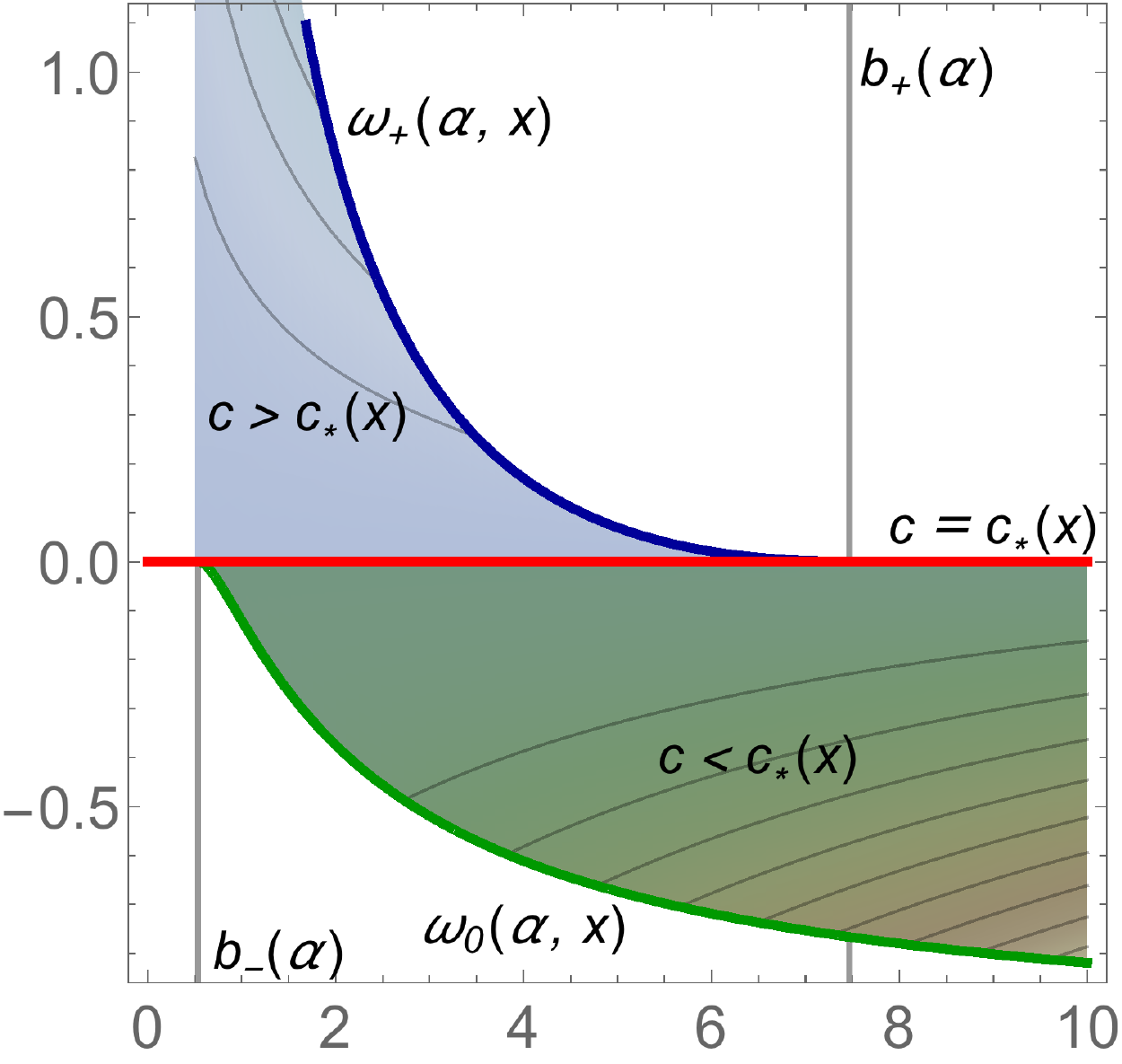}
\caption{Density plot of the rate function in the $(x,\omega)-$plane and the $(x,c)-$plane for $\alpha=2$}
\label{Ratevsc}
\end{center}
\end{figure*}
As shown in \cite{Perez2014b,Perez2014c} one can obtain an exact expression of the action \eref{eq:af} in terms of elliptic integrals (see  \ref{app:A}). In  \Fref{Ratevsc} we present a density plot of the expression \eref{eq:af} in terms of the pair of variables $(x,c)$.\\
Finally, we can extract the rate functions of the smallest and largest eigenvalues from $\Psi(c,x)$.  Let us denote as $\Psi^{(\pm)}_{{\rm M}} (x)$ and $\Psi^{(\pm)}_{{\rm m}} (x)$ the left (minus sign) and right (plus sign) rate functions of the largest and smallest eigenvalue, respectively.  Indeed, it can be shown that
\beeq{
\Psi^{(-)}_{m}(x)=\lim_{c\to 0^{+}}\frac{\Psi(c,x)}{c}\,,\quad\quad x\leq b_{-}(\alpha)\,,\label{f1}\\
\Psi^{(+)}_{m}(x)=\lim_{c\to 0^{+}}\Psi(c,x)\,,\quad\quad x\geq b_{-}(\alpha)\,,\label{f2}
}
and
\beeq{
\Psi^{(-)}_{M}(x)=\lim_{c\to 1^{-}}\Psi(c,x)\,,\quad\quad x\leq b_{+}(\alpha)\,,\label{f3}\\
\Psi^{(+)}_{M}(x)=\lim_{c\to 1^{-}}\frac{\Psi(c,x)}{1-c}\,,\quad\quad x\geq b_{+}(\alpha)\,.\label{f4}
}
where the expressions of $\Psi^{(\pm)}_{{\rm M}} (x)$ and $\Psi^{(\pm)}_{{\rm m}} (x)$ correspond to those reported in \cite{Katzav2010,Majumdar2009,Vivo2007}. In \Fref{RFs} we present a comparison  of the left hand side and right hand side of Equations (\ref{f1}-\ref{f4}).
\begin{figure*}[t]
\begin{center}
\includegraphics[width=7cm,height=5cm]{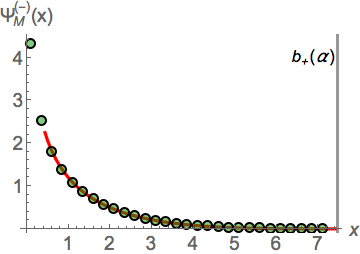}\quad\includegraphics[width=7cm,height=5cm]{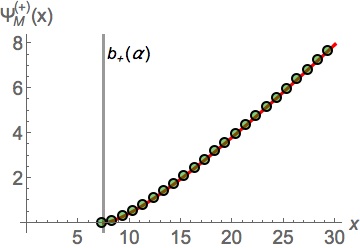}\\
\includegraphics[width=7cm,height=5cm]{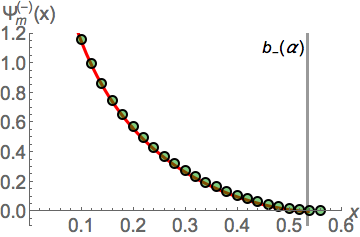}\quad\includegraphics[width=7cm,height=5cm]{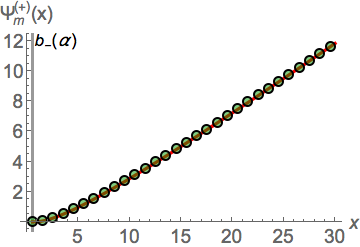}
\caption{Comparison between $\Psi(c,x)$ for the extreme values (green filled circles) according to the formulas (\ref{f1}-\ref{f4}) and the results obtained in  \cite{Katzav2010,Majumdar2009,Vivo2007} (solid red lines). The plots have been done by choosing $\alpha=2$. }
\end{center}
\label{RFs}
\end{figure*}

\section{Monte Carlo Simulations}
\label{sec:MonteCarlo}
\begin{figure*}[t]
\includegraphics[width=7cm,height=5cm]{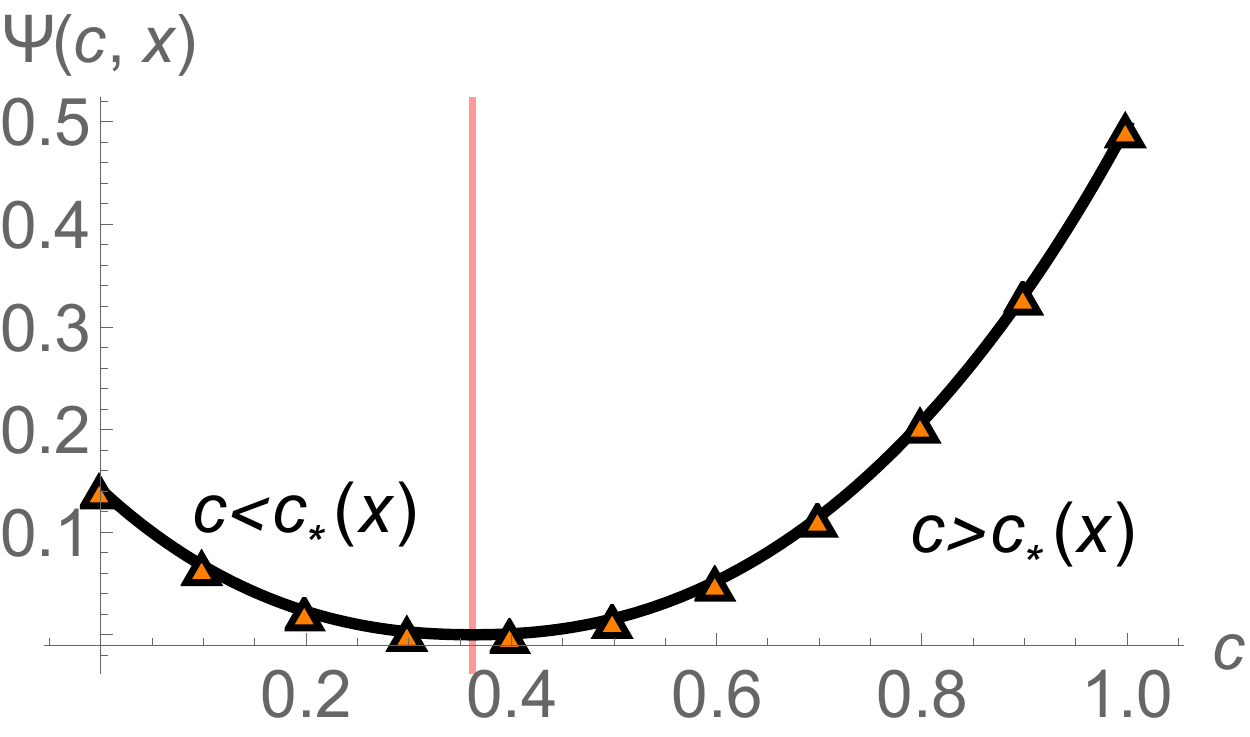}\quad\quad\includegraphics[width=7cm,height=5cm]{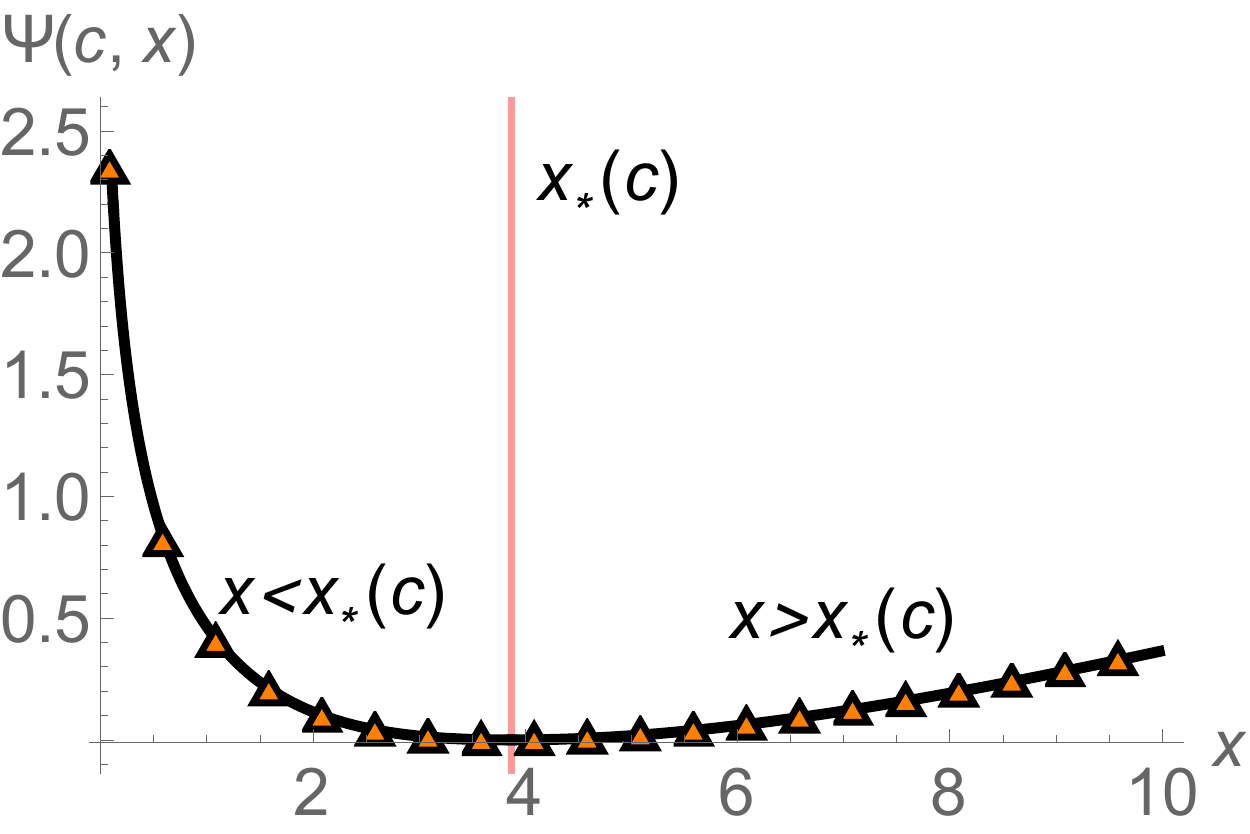}
\caption{Left: Rate function vs $c$ for $\alpha=2$ and $x=2$ and comparison with simulations with $N=500$, with time window for relaxation of 10000 steps and average over a window of 10000 MC steps. Right: Rate function vs $x$ for $c=7/10$ and comparison with Monte Carlo simulations. The latter have been performed with $N=500$ and $N_{{\rm left}}=350$ and with the same time windows as before for thermalising and averaging}
\label{RatevscMC}
\end{figure*}

We have compared our results with Monte Carlo simulations of the Coulomb fluid by  first equilibrating it using Metropolis algorithm and then generating equilibrium samples to estimate averages. If we denote $\bracket{\cdots}_{{\rm MC}}$ as averages over the Monte Carlo Markov Chain, then the action can be estimated as: 
\beeq{
\mathcal{A}_0(c,x)=\frac{1}{N^2}\left[\bracket{F(\bm{y})}_{{\rm MC}}+uN\ln N+N(N-1)\ln N\right]\,.
}
In  \Fref{RatevscMC}, we show a comparison between the exact result (solid lines) together the estimates resulting from Monte Carlo simulations (orange filled  triangles).
The plot on the right panel corresponds to fixing a value of $c$. From the point of view of the Monte Carlo simulation this is achieved by putting a number $N_{{\rm left}}$ of eigenvalues to the left of the barrier at $x$ out of a total $N$ such that $c= N_{{\rm left}}/N$. Then the Metropolis algorithm is implemented as normally but eigenvalues are  not allowed to cross the barrier. This is precisely the method that was used to generate the histograms of the deformed MP law appearing in \Fref{fig:dmpls}.

\section{Conclusions}
\label{sec:Conclusions}
In this work we have presented a unified way to obtain the large deviations of extreme and bulk eigenvalues of the Wishart-Laguerre ensemble. Using the fluid Coulomb picture we are able to derive a rate function $\Psi(c,x)$ depending on two variables: the fraction of eigenvalues $c$ to the left of the barrier at position $x$. It was noted that if we fix $x$ and vary $c$, $\Psi(c,x)$ gives information about the large deviations of the SIN. On the other hand, if we fix $c=k/N$ and vary $x$ we obtain the large deviations of the $k$-th eigenvalue, particularly when $k=1$ for the smallest eigenvalue\cite{Katzav2010} and $k=N$ for the largest\cite{Vivo2007,Majumdar2009}. We have contrasted our results with Monte Carlo simulations showing perfect agreement.\\
It would be interesting to see how this method applies and generalises the work done in the Jacobi ensemble (see for instance \cite{Ramli2012}). This line of inquiry is currently under way.

\ack
This work has been funded by the program UNAM-DGAPA-PAPIIT IA101815.

\appendix
\section{Action in terms of elliptic integrals}
\label{app:A}
We first recall all the previous results for the integrals of the resolvent in terms of elliptic integrals\footnote{For this the book of \cite{Byrd1971} is extremely useful.}. We divide them into the two cases. For $c>c_{\star}$  we have that
\beeq{
\mathcal{A}_{0}(c,x)&=\frac{1+\alpha-x\omega}{2}+\frac{\alpha}{4}\left(\alpha\ln(\lambda_{-})+\tilde{I}_1(\lambda_{+},\lambda_0,x,\lambda_{-})\right)\nonumber\\
&+\frac{1}{2}\left(c+\frac{\alpha}{2}\right)K_1(\lambda_+,\lambda_0,x,\lambda_{-})\nonumber \\
&-\frac{1}{2}\left(1+\frac{\alpha}{2}\right)\left(-\lambda_{+}+(2+\alpha)\ln(\lambda_{+})+\tilde{I}(\lambda_{+},\lambda_0,x,\lambda_{-})\right)\nonumber\,,
}
with the following definitions
\beeq{
\hspace{-2cm}\tilde{I}_1(a,b,c,d)&=\frac{(a+2b-c)(c-d) }{ \sqrt{(a-c) (b-d)}}F\left(\theta_1,k\right)+\sqrt{(a-c) (b-d)} E\left(\theta_1,k\right)\nonumber\\
\hspace{-2cm}&+\frac{(d-c) (a+b-c+d) }{\sqrt{(a-c) (b-d)}}\Pi \left(\theta_1,\tilde{\alpha}^2,k\right)\nonumber\\
\hspace{-2cm}&+\frac{(a-d) (d-b)}{\sqrt{(a-c) (b-d)}}\frac{ {\rm cn}\left(F\left(\theta_1,k\right),k\right) {\rm dn}\left(F\left(\theta_1,k\right),k\right) {\rm sn}\left(F\left(\theta_1,k\right),k\right)}{1-\tilde{\alpha}^2 {\rm sn}^2\left(F\left(\theta_1,k\right),k\right)}\nonumber\\
\hspace{-2cm}&-\frac{2 a b (c-d) }{c \sqrt{(a-c) (b-d)}}\Pi\left(\theta_1,\frac{d(b-c)}{c(b-d)},k\right)+\sqrt{\frac{abd}{c}}\ln\frac{4 a b c }{a b c+a b d-a c d-b c d}\nonumber\,,\\
\hspace{-2cm}K_1(a,b,c,d)&=\frac{1}{c \sqrt{(a-c) (b-d)}}\Bigg[c (c-a) (b-d) E\left(k\right)\nonumber\\
\hspace{-2cm}&-(c-d) \Bigg(c (b-d) K\left(k\right)+c (a+b-c+d) \Pi \left(\frac{b-c}{b-d},k\right)\nonumber\\
\hspace{-2cm}&-2 a b \Pi \left(\frac{(b-c) d}{c (b-d)},k\right)\Bigg)\Bigg]\nonumber\,,\\
\hspace{-2cm}\tilde{I}(a,b,c,d)&=-\frac{1}{2\sqrt{(a-c)(b-d)}}\Bigg[2 (a-b) (a-c+2 d) F\left(\theta_2,k\right)\nonumber\\
\hspace{-2cm}&+2 (a-c) (b-d) E\left(\theta_2,k\right)-2 (a-b) (a+b-c+d) \Pi \left(\theta_2,\frac{b-c}{a-c},k\right)\nonumber\\
\hspace{-2cm}&+4 d (b-a) \Pi \left(\theta_2,\frac{b (a-d)}{a (b-d)},k\right)\nonumber\\
\hspace{-2cm}&+\sqrt{(a-c) (b-d)} \left((a+b-c+d) \ln \left(\frac{4 a}{a+b-c-d}\right)-a-b+c+d\right)\Bigg]\nonumber\,,\\
\hspace{-2cm}&\theta_1=\sin ^{-1}\sqrt{\frac{(a-c) d}{(a-d)c }}\,,\quad k^2=\frac{(b-c)(a-d)}{(a-c)(b-d)}\,,\quad\quad\theta_2=\sin ^{-1}\left(\sqrt{\frac{b-d}{a-d}}\right)\nonumber\,.
}
The expression for the action for the case $c<c_{\star}$ is almost similar to the previous one. We just need to replace the function  $K_1\to-K_2$ and write
\beeq{
\mathcal{A}_{0}(c,x)&=\frac{1+\alpha-x\omega}{2}+\frac{\alpha}{4}\left(\alpha\ln(\lambda_{-})+\tilde{I}_1(\lambda_{+},\lambda_0,x,\lambda_{-})\right)\nonumber\\
&-\frac{1}{2}\left(c+\frac{\alpha}{2}\right)K_2(\lambda_+,x,\lambda_0,\lambda_{-})\nonumber \\
&-\frac{1}{2}\left(1+\frac{\alpha}{2}\right)\left(-\lambda_{+}+(2+\alpha)\ln(\lambda_{+})+\tilde{I}(\lambda_{+},\lambda_0,x,\lambda_{-})\right)\nonumber
}
with
\beeq{
\hspace{-2cm}K_2(a,b,c,d)&=\frac{1}{\sqrt{(a-c) (b-d)}}\Bigg[(c-d) \left((b-d) K\left(k\right)+(a-b+c+d) \Pi \left(\frac{b-c}{b-d},k\right)\right.\nonumber\\
\hspace{-2cm}&\left.-2 a \Pi \left(\frac{(b-c) d}{c (b-d)}, k\right)\right)+(a-c) (b-d) E\left(k\right)\Bigg]\,,\quad k^2=\frac{(b-c) (a-d)}{(a-c) (b-d)}\nonumber
}

\end{document}